# Origin of the giant spin Hall effect in BiSb topological insulator

Takanori Shirokura[1], Kenichiro Yao[1], Yugo Ueda[1], and Pham Nam Hai[1,2,3*]

[1]*Department of Electrical and Electronic Engineering, Tokyo Institute of Technology,*

*2-12-1 Ookayama, Meguro, Tokyo 152-0033, Japan*

[2]*Center for Spintronics Research Network (CSRN), The University of Tokyo,*

*7-3-1 Hongo, Bunkyo, Tokyo 113-8656, Japan*

[3]*CREST, Japan Science and Technology Agency,*

*4-1-8 Honcho, Kawaguchi, Saitama 332-0012, Japan*

*Corresponding author: pham.n.ab@m.titech.ac.jp

**The giant spin Hall effect (SHE) at room temperature is one of the most attractive feature of topological insulators (TIs) for applications to nano-scale spin devices. Its origin, however, remains a controversial problem. Here, we identify the origin of the giant SHE in BiSb thin films by measuring the spin Hall angle $\theta_{SH}$ under controllable contribution of surface and bulk conduction. We found that $\theta_{SH}$ of a $Bi_{0.6}Sb_{0.4}$ TI thin film takes colossal values (450 ~ 530 at 8 K, and 38 at 300 K), and is almost governed by contribution from the topological surface states. Meanwhile, $\theta_{SH}$ in a $Bi_{0.2}Sb_{0.8}$ semi-metallic thin film without topological surface states drastically decreases. Our results provide a quantitative tool for analysing the origin of the giant SHE in TI thin films, as well as a strategy for designing spin current source utilizing the surface states of TI in high-performance nano-scale spin devices.**

The discovery of giant SHE with $\theta_{SH}$ exceeding unity at room temperature in several TI ultrathin films has attracted much attention for possible applications to nano spin devices, such as spin-orbit-torque magneto resistive random access memories and spin-torque oscillators.[1-4] An early experiment on the giant SHE in the well-studied $Bi_2Se_3$ TI assumed the surface state origin of the observed $\theta_{SH}$ (2 – 3.5), even though the Fermi level in $Bi_2Se_3$ lies in the conduction band and the contribution from the bulk conduction cannot be ignored.[2] The parallel conduction in the metallic surface and degenerated bulk states in $Bi_2Se_3$ cannot be quantitatively separated, and leads to large discrepancies in different measurements of $\theta_{SH}$ and even contradicting conclusion on the origin of the giant SHE in $Bi_2Se_3$.[5,6] More recently, systematic measurements of SHE in $(Bi_{1-x},Sb_x)_2Te_3$ TI thin films, whose Fermi level can be tuned to the band gap, show that SHE is almost twice larger when the Fermi level is in the valence band (bulk states dominant) than that when the Fermi level is in the band gap (topological surface states dominant), adding more controversy to the problem.[7] From both fundamental and technological aspects, the origin of the giant SHE is the must solve problem for designing any nano spin devices using TI as the pure spin current source.

In this letter, we identify the origin of the giant SHE in BiSb thin films, by measuring the spin Hall angle $\theta_{SH}$ under controllable contribution of surface and bulk conduction. $Bi_{1-x}Sb_x$ ($0.07 < x < 0.22$) is the first three dimensional TI whose topological surface states have been detected using angle-resolved photoemission spectroscopy (ARPES)[8,9], scanning tunneling spectroscopy[10], and quantum transport measurements.[11,12] Recently, we observed that 10 nm-thick $Bi_{0.9}Sb_{0.1}$ thin films with pseudo-cubic (012) surface show the highest room-temperature spin Hall conductivity ($1.3\times10^7$ $\frac{\hbar}{2e}\Omega^{-1}m^{-1}$) among all known materials thanks to the very large $\theta_{SH}$ ~ 52 and high electrical conductivity $\sigma$~ $2.5\times10^5$ $\Omega^{-1}m^{-1}$, making $Bi_{1-x}Sb_x$ a very promising candidate for the pure spin current source in nano-scale spin devices.[1] In terms of fundamental properties, $Bi_{1-x}Sb_x$ has two unique features compared with other TIs. First, it has long Fermi wavelength (~ 40 nm) and much higher bulk mobility (~ $10^4$ $cm^2V^{-1}s^{-1}$) than other TIs, so that quantum confinement can occur in BiSb thin films even at room temperature. As a result, the band gap of BiSb thin films becomes larger at reduced thickness, and BiSb can become insulating in the region where it is semimetal in the bulk. For $Bi_{1-x}Sb_x$ thin films as thick as 90 nm, quantum confinement extends the insulating region from $x$=0% to at least 35%.[13] Secondly, the Fermi level in BiSb is always in the band gap. Thus, BiSb thin films always have metallic surface states and insulating bulk states.[13,14] This makes it possible to quantitatively evaluate the contribution of the surface states and bulk states to the giant SHE. Let us consider the electrical transport in a BiSb thin film for a general case shown in the inset of Fig. 1(a), where $\sigma_S$ and $\sigma_B$ are the conductivity of the surface and bulk states, $t_S$ is the total penetration depth of the upper and lower surface states, $t_B$ is the thickness of the bulk states, $\theta_{SH}^{S}$ and $\theta_{SH}^{B}$ are the spin Hall angle of the surface and bulk states, respectively. Because the Fermi level is in the band gap, the bulk conductivity follows $\sigma_B$ = $\sigma_0\exp(-\frac{E_g}{2k_BT})$, while $\sigma_S$ is nearly temperature-independent. The temperature dependence of the total conductivity is given by $\sigma(T)=\frac{\sigma_S t_S}{t_B}+\sigma_0\exp(-\frac{E_g}{2k_BT})$ (Eq. 1). Therefore, for $Bi_{1-x}Sb_x$ thin films with appropriate thickness and band gap, we can quantitatively evaluate the contribution of the surface states to the total conductivity $\Gamma \equiv \frac{\sigma_S t_S}{\sigma_S t_S + \sigma_B t_B}$ by investigating the temperature dependence of the electrical resistivity, which is indispensable for identifying the origin of the giant SHE.

As a demonstration, we first show in Fig. 1(a) the temperature dependence of the normalised resistivity of $Bi_{0.89}Sb_{0.11}$ thin films with thickness of 10 nm (green dotted line), 41 nm (red dotted line), and 92 nm (blue dotted line), respectively. The solid lines are fitting curves using Eq. (1), which show good agreement with the experimental data, yielding Γ(300 K) = 20.2 %, $E_g$= 34.2 meV for the 92 nm-thick film, Γ(300 K) = 38.3 %, $E_g$ = 66.6 meV for the 41 nm-thick film, and Γ(8 - 300 K) = 100 % for the 10 nm-thick film. One can see that the bulk conduction is dominant for the 92 nm-thick film. As the thickness is reduced to 41 nm, the surface state contribution began to rise significantly due to the increasing band gap. When the thickness is furthered reduced to 10 nm, the band gap becomes large enough so that metallic surface conduction is dominant. Thus, BiSb with intermediate thickness (40 – 50 nm) has tunable surface and bulk conduction, and is suitable for studying the origin of the giant SHE. In this work, we prepare two samples; one is MnAs (5 nm) / $Bi_{0.6}Sb_{0.4}$(001) (50 nm) (sample A) and the other is MnAs (3.2 nm) / $Bi_{0.2}Sb_{0.8}$(001) (50 nm) (sample B), grown on GaAs(111)A substrates by molecular beam epitaxy (MBE), whose schematic cross section are shown in Fig. 1(b). Here, the hexagonal coordinate indexing and the cubic coordinate indexing are used for BiSb and GaAs, respectively. The temperature dependence of the resistivity reveals that the $Bi_{0.6}Sb_{0.4}$ (50 nm) layer behaves as a topological insulator with a band gap of 33.2 meV and mixing of surface and bulk conduction, and that Γ changes from 60 % to 100 % when temperature decreases from 300 K to 8 K. On the other hand, the $Bi_{0.2}Sb_{0.8}$ (50 nm) layer behaves purely as a semimetal layer (see Supplementary Note 1) .[13] The MnAs top layers have in-plane magnetisation for detection of the injected spin current. These samples are patterned into Hall bars with dimension of 50 μm × 100-200 μm by Ar ion milling and photolithography for transport measurements. To study the spin Hall effect of BiSb, we use the direct current (DC) planar Hall technique with an in-plane rotating magnetic field [3,15,16] over a wide range of temperature from 8 K to 300 K. Figure 1(c) shows the schematic experimental setup and the coordinate system of the DC planar Hall method. A DC current ***I*** is injected to the bi-layer along the *x*-direction. The Hall resistance $R_H$ is measured under a constant magnetic field $H_{\mathrm{ext}}$ = 5 kOe rotating in the *xy*-plane for $\theta = 0$ - 360° with respect to the *x* axis. The SHE from the BiSb layer generates a transverse field-like effective field $\boldsymbol{H}_{\mathrm{T}}$ and a perpendicular anti-damping-like effective field $\boldsymbol{H}_{\mathrm{SO}}$ acting on the MnAs layer. Here, $\boldsymbol{H}_{\mathrm{SO}}$ is given by $\boldsymbol{H}_{\mathrm{SO}} = \frac{\hbar\theta_{\mathrm{SH}}J_{\mathrm{C}}}{2eM_{\mathrm{S}}t_{\mathrm{FM}}}\left(-\hat{\boldsymbol{\sigma}}\times\hat{\boldsymbol{m}}\right) = -\frac{\hbar\theta_{\mathrm{SH}}J_{\mathrm{C}}}{2eM_{\mathrm{S}}t_{\mathrm{FM}}}(\cos\phi)\hat{\boldsymbol{z}}$, where $\hbar$ is the Dirac's constant, $J_{\mathrm{C}}$ is the charge current density in the BiSb layer, $e$ is the electronic charge, $M_{\mathrm{S}}$ is the saturation magnetisation of the MnAs layer, $t_{\mathrm{FM}}$ is the thickness of the MnAs, $\hat{\boldsymbol{\sigma}}$ is the spin polarization unit vector, and $\hat{\boldsymbol{m}}$ is the magnetisation unit vector. $\phi$ is the magnetisation direction in the *xy*-plane with respect to the *x*-axis, and given by $\phi\,(\pm I) = \mathrm{Tan}^{-1}\{(\sin\theta \pm \delta)/\cos\theta\}$, where $\delta = H_{\mathrm{T}}/H_{\mathrm{ext}}$. Since $H_{\mathrm{SO}}$ generates a z-component of the magnetisation, the Hall resistance $R_{\mathrm{H}}$ is given by

$$R_{\mathrm{H}}(\pm I) = R_{\mathrm{PHE}}\sin 2\phi + R_{\mathrm{AHE}}^{\mathrm{SO}}\cos\phi + R_{\mathrm{AHE}}^{x}\cos\phi + R_{\mathrm{AHE}}^{y}\sin\phi + C \qquad (2)$$

The first term comes from the planar Hall effect (PHE). The second term comes from the anomalous Hall effect (AHE) due to the $H_{\mathrm{SO}}$-induced z-component of the magnetisation, and the third and fourth term arise from small mis-alignment of the $H_{\mathrm{ext}}$ rotating surface from the $xy$-plane. The fifth term is the experimental Ohmic offset.

Figure 2 shows the $\theta$-dependence of $R_{\mathrm{H}}$ as a function of $I$ in sample A at 8 K and those in sample B at 10 K. The blue circles show the measurement data and the red solid lines show the fitting results using Eq. (2) with $R_{\mathrm{PHE}}$, $R^{\mathrm{SO}}_{\mathrm{AHE}} + R^{x}_{\mathrm{AHE}}$, $R^{y}_{\mathrm{AHE}}$, $C$ and $\delta$ as fitting parameters. The fitting curves agree very well with the experimental data. $R_{\mathrm{H}}$ of sample A shows strong dependence on the input current, as shown in Fig. 2(a) – 2(f). In contrast, much weaker input current dependence is observed for sample B, as shown in Fig. 2(g) – 2(l). Hence, $H_{\mathrm{SO}}$ in sample A is much stronger than that in sample B. Because $R^{\mathrm{SO}}_{\mathrm{AHE}}$ depends on $J_{\mathrm{C}}$ while $R^{x}_{\mathrm{AHE}}$ does not, we can extract $R^{\mathrm{SO}}_{\mathrm{AHE}}$ as a function of $J_{\mathrm{C}}$. Figure 3 shows the $J_{\mathrm{C}}$-dependence of $R^{\mathrm{SO}}_{\mathrm{AHE}}$ of sample A (Fig. 3(a) – 3(h)) and sample B (Fig. 3(i) – 3(l)) as a function of temperature, respectively. Here, for calculation of $J_{\mathrm{C}}$, we calibrated the conductivity of the MnAs layer from the amplitude of PHE ($R_{\mathrm{PHE}}$) (see Supplementary Note 2). The blue dots and the blue solid lines show the measurement data and linear fitting of $R^{\mathrm{SO}}_{\mathrm{AHE}}$, respectively. The red solid lines show $H_{\mathrm{SO}}$ calculated from $R^{\mathrm{SO}}_{\mathrm{AHE}} \frac{dH_{\mathrm{perp}}}{dR_{\mathrm{AHE}}} \frac{H_{\mathrm{ext}} + H_{\mathrm{deg}}}{H_{\mathrm{deg}}}$, where $\frac{dR_{\mathrm{AHE}}}{dH_{\mathrm{perp}}}$ is the gradient of the anomalous Hall resistance vs. perpendicular magnetic field (see Supplementary Note 3) , and $\frac{H_{\mathrm{ext}} + H_{\mathrm{deg}}}{H_{\mathrm{deg}}}$ is the correction factor for the large in-plane $H_{\mathrm{ext}}$ = 5 kOe and the small demagnetising field $H_{\mathrm{deg}} = 4\pi M_{\mathrm{s}}$ = 5 kOe of the MnAs layers. Here, we took into account the shunt circuit effect of the bias current, the short circuit effect of the Hall voltage, and the normal Hall effect in estimating $\frac{dR_{\mathrm{AHE}}}{dH_{\mathrm{perp}}}$ (Supplementary Note 3).[17] In Fig. 3(a) – 3(h), the sign of $H_{\mathrm{SO}}$ in sample A is the same as that of Pt,[15,18] and agrees with previous reports on Bi-based TIs.[1-3,7] However, the sign of $H_{\mathrm{SO}}$ in sample B is reversed, demonstrating the fundamental difference in the origin of $H_{\mathrm{SO}}$ between sample A (TI) and sample B (semimetal).

Figure 4 (a) shows the temperature dependence of $\theta_{\mathrm{SH}}$ in sample A (blue dots) and sample B (red dots) calculated by $\theta_{\mathrm{SH}} = \frac{2eM_{\mathrm{S}}t_{\mathrm{FM}}H_{\mathrm{SO}}}{\hbar J_{\mathrm{C}}}$. For a reference, we also show the room-temperature $\theta_{\mathrm{SH}}$ (=52) of a 10 nm-thick BiSb(012) layer (green dot in Fig. 4(a)).[1] Note that $\theta_{\mathrm{SH}}$ calculated in this way is the nominal value of the whole BiSb layers with contribution from both the surface and the bulk states. $\theta_{\mathrm{SH}}$ of sample A dramatically increases from 38 at 300 K to a colossal value of 450 at 8 K, and is significantly larger than that of sample B (-4.4 at 10 K and -0.98 at 154 K). The much higher $\theta_{\mathrm{SH}}$ in sample A is an evidence for the important contribution of surface states to the giant SHE in BiSb TI thin films.

The colossal value of $\theta_{\mathrm{SH}}$ = 450 at 8 K corresponds to the very large spin Hall conductivity of $\sigma_{\mathrm{SH}} \sim 1.1\times10^{8}\,\frac{\hbar}{2e}\Omega^{-1}\mathrm{m}^{-1}$ for BiSb, which is even higher than the "charge" conductivity of silver, the most conductive metal ($\sigma \sim 6.3\times10^{7}\ \Omega^{-1}\mathrm{m}^{-1}$). To check that we do not overestimate $\theta_{\mathrm{SH}}$ at low temperatures due to thermoelectric effects, such as the anomalous Nernst effect in the MnAs layer, we independently estimate $\theta_{\mathrm{SH}}$ from the amplitude of the planar Hall resistance $R_{\mathrm{PHE}}$ as a function of the bias current (see Supplementary Note 2). Our analysis shows that $R_{\mathrm{PHE}}$ should be reduced by $\frac{\Delta R_{\mathrm{PHE}}}{R_{\mathrm{PHE}}} = -\frac{(H_{\mathrm{SO}})^2}{2(H_{\mathrm{ext}}+H_{\mathrm{deg}})^2} = -\frac{(\alpha J)^2}{2(H_{\mathrm{ext}}+H_{\mathrm{deg}})^2}$ (see Eq. S6 in the Supplementary Note 2). Here, $\alpha$ is the proportional constant between $H_{SO}$ and $I$, and directly related to $\theta_{SH}$. In contrast, any thermoelectric effects would cause a linear dependence of $R_{PHE}$ on $I$. Our data show a clear quadratic and no linear dependence of $R_{PHE}$ on $I$ for the sample A (Fig. S2 in Supplementary Note 2, also see Supplementary Note 4). Thus, thermoelectric effects are absent or negligible in our bi-layers. Furthermore, by fitting to the $\frac{\Delta R_{PHE}}{R_{PHE}}$ - $I$ data, we can independently estimate $\alpha$ and $\theta_{SH}$. Our data at 8 K yields $\alpha$ = 0.22 kOe/mA and $\theta_{SH}$ = 530, which is even higher than $\theta_{SH}$ = 450 estimated from the $R_{AHE}^{SO}$. The 15% difference between the two methods may be due to uncertainty in estimation of the $\frac{dR_{AHE}}{dH_{perp}}$ gradient used in the former method. Our results double check the colossal $\theta_{SH}$ of BiSb at low temperature.

Next, we quantitatively evaluate the contribution of the surface states to the nominal $\theta_{SH}$. Assuming the parallel conduction model in the inset of Fig. 1(a) with $t_B >> t_s$, $\theta_{SH}$ for a TI layer is given by (Supplementary Note 5),

$$\theta_{SH} = \Gamma \frac{t_B}{t_S} \theta_{SH}^{S} + \frac{2e}{\hbar} \frac{\sigma_{SH}^{B} t_B}{\sigma_S t_S + \sigma_B t_B}, \quad (3)$$

where $\sigma_{SH}^{B} = \frac{\hbar}{2e} \theta_{SH}^{B} \sigma_B$ is the spin Hall conductivity of bulk states. Eq. (3) explains why there are large discrepancies in different measurements of $\theta_{SH}$ for the case of $Bi_2Se_3$ with different thickness and bulk conductivity. Even if we assume $\sigma_S$ and $t_S$ are intrinsic parameters of a TI and do not change with the growth condition, $\theta_{SH}$ can still vary with $\sigma_B, t_B$, and $\Gamma$. In the limit of $\sigma_S t_S << \sigma_B t_B$ (bulk conduction is dominant), $\theta_{SH}$ is equal to $\theta_{SH}^{B}$, which is typically of order of 0.1. In the opposite case of $\sigma_S t_S >> \sigma_B t_B$, $\theta_{SH}$ approaches $\frac{t_B}{t_S} \theta_{SH}^{S}$ but still depends on the ratio $\frac{t_B}{t_S}$. In general, in order to understand the origin of SHE in TI, it is necessary to know the values of the parameters in Eq. (3). In our case of BiSb, $\sigma_S t_S$ and $\sigma_B t_B$ can be experimentally deduced from the temperature dependence of the total conductivity and fitting to Eq. (1). However, the exact value of $t_s$ is unknown. Thus, to evaluate the surface and the bulk contribution, we introduce the nominal "sheet spin Hall angle" $q_{SH} \equiv \frac{\theta_{SH}}{t_B + t_S} \approx \frac{\theta_{SH}}{t_B}$ for the whole layer, and the "surface sheet spin Hall angle" $q_{SH}^S \equiv \frac{\theta_{SH}^S}{t_S}$ for the surface states. Substituting $q_{SH}$ and $q_{SH}^S$ to Eq. (3), we obtain $q_{SH}^S = \frac{1}{\Gamma} q_{SH} - \frac{2e}{\hbar} \frac{\sigma_{SH}^B}{\sigma_S t_S}$ (4), which reflects the intrinsic SHE of the surface states. Using this equation, we can estimate $q_{SH}^S$ using the experimental values of Γ, $q_{SH}$, $\sigma_S t_S$, and the calculated value of $\sigma_{SH}^B$ ~ $-9.3\times10^4 \frac{\hbar}{2e} \Omega^{-1}m^{-1}$ from the first principle calculation.[19] Figure 4 (b) shows the values of $q_{SH}^S$ (blue dots) and $q_{SH}$ (blue squares) of sample A as a function of temperature. $q_{SH}^S$ rapidly increases from 1.3 nm$^{-1}$ at room temperature to about 9 nm$^{-1}$ at temperatures below 100 K. The inset in Fig. 4 (b) shows the contribution of the bulk states to the nominal sheet spin Hall angle, $R \equiv \frac{2e}{\hbar} \frac{|\sigma_{SH}^B|}{q_{SH}(\sigma_S t_S + \sigma_B t_B)}$, which is always lower than 1%. Our results clearly show that the giant SHE in BiSb TI thin film originates from the surface states, even in the $Bi_{0.6}Sb_{0.4}$ layer as thick as 50 nm. Note that although the room-temperature $q_{SH}^S$ = 1.3 nm$^{-1}$ of the 50 nm-thick $Bi_{0.4}Sb_{0.6}$(001) layer is already large, it is still smaller than the room-temperature value ($q_{SH}^S$ = 5.2 nm$^{-1}$) of a 10 nm-thick BiSb(012) layer (green dot in Fig. 4(b)).[1] The difference may come from the fact that there are three Dirac cones on the pseudo-cubic surface of BiSb(012), which were theoretically predicted [20] and experimentally confirmed by ARPES observation,[21] comparing with one Dirac cone on the hexagonal surface of BiSb(001). Since Dirac cones on the surface of TI are monopoles of Berry flux with the same chirality, more Dirac cones mean larger total Berry flux and thus higher surface spin Hall conductivity. The correlation between the sheet spin Hall angle and the numbers of Dirac cones on the BiSb surfaces gives another evidence for the surface state origin of the giant SHE in BiSb.

In summary, by quantitatively evaluating the SHE under controllable contribution of surface and bulk conduction at various temperature and surface orientation, we have shown that the giant SHE in BiSb is entirely governed by the surface states from 8 K up to room temperature. Our results provide a quantitative tool for analysing the origin of the giant SHE in TI thin films, as well as a strategy for designing spin current source utilizing the surface states of TI in high-performance nano-scale spin devices.

**Acknowledgement**

This work is supported by Grant-in-Aid for Challenging Exploratory Research (No. 16K14228), Nanotechnology platform 12025014 (F-17-IT-0011) from MEXT, and JST CREST (JPMJCR18T5). Data of sample B are obtained with supports from TDK Corporation. We thank the Material Analysis Division and Laboratory for Future Interdisciplinary Research of Science and Technology at Tokyo Institute of Technology, and M. Tanaka Laboratory at University of Tokyo for their technical supports.

## Methods

**MBE growth.** The thin films were grown on semi-insulating GaAs(111)A substrates by using ultrahigh vacuum MBE system. After the surface oxide layer of the GaAs substrate was removed by annealing at 580°C, a GaAs buffer layer was grown to obtain an atomically smooth surface. Then, a 50 nm-thick $Bi_{0.6}Sb_{0.4}$(001) or $Bi_{0.2}Sb_{0.8}$(001) layer was grown at a substrate thermocouple temperature of 200°C. Finally, a 5 nm (or 3.2 nm)-thick MnAs(001) layer was grown on top of the BiSb layer at a substrate thermocouple temperature of 320°C as the ferromagnetic layer. A 23.3 nm–thick MnAs(001) single layer was also grown on a GaAs(111)B substrate as a reference.

**Hall bar fabrication.** The samples were patterned into Hall bar structures by standard optical lithography and Ar ion milling.

**SOT measurements.** A Keithley 2400 Sourcemeter was used as the current source for DC planar Hall measurements. The Hall voltage was measured using a Keithley 2002 Multimeter. The Hall bars were mounted inside a Gifford-McMahon cryostat equipped with a computer-controlled rotatable electromagnet.

## Author Contributions

T.S. and K.Y. (equal contribution) grew the MnAs/BiSb bi-layers, fabricated the Hall devices, and performed Hall effect measurements; Y. U. evaluated the conductivity of various BiSb layers; P.N.H. planned the experiments; T.S., K.Y, P.N.H. analysed the data. T.S. and P.N.H wrote the manuscript, with comments from K.Y. and Y. U.

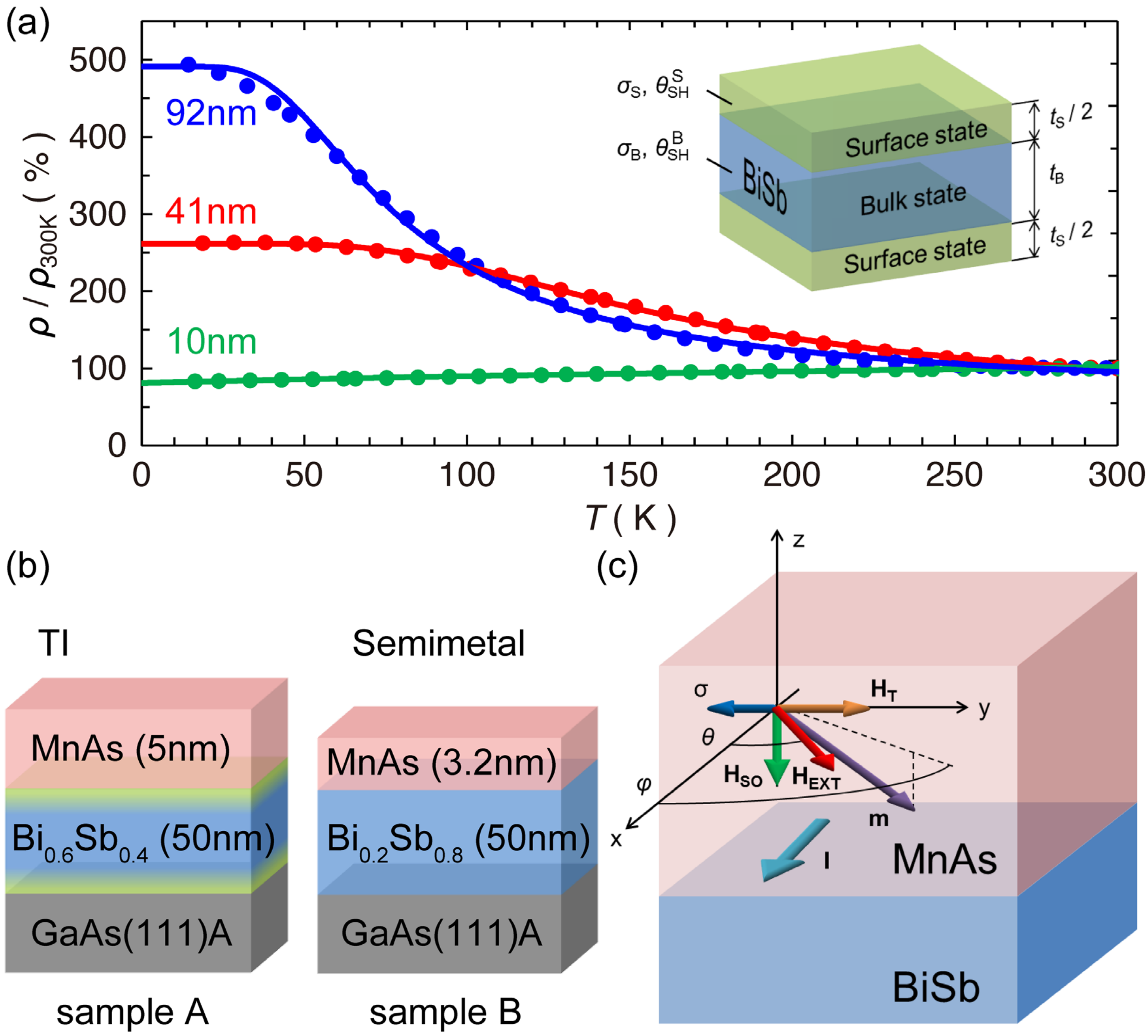

**FIG. 1. (a)** Temperature dependence of the normalised resistivity in $Bi_{0.89}Sb_{0.11}$ thin films with thickness of 10 nm (green), 41 nm (red) and 92 nm (blue). Dots and solid lines show the experimental data and fitting to a parallel conduction model of BiSb (inset). **(b)** Schematic structure of a MnAs / BiSb (TI) bi-layer (sample A) and a MnAs / BiSb (semimetal) bi-layer (sample B). **(c)** Schematic illustration of the experiment setup and coordination system for DC planar Hall measurements.

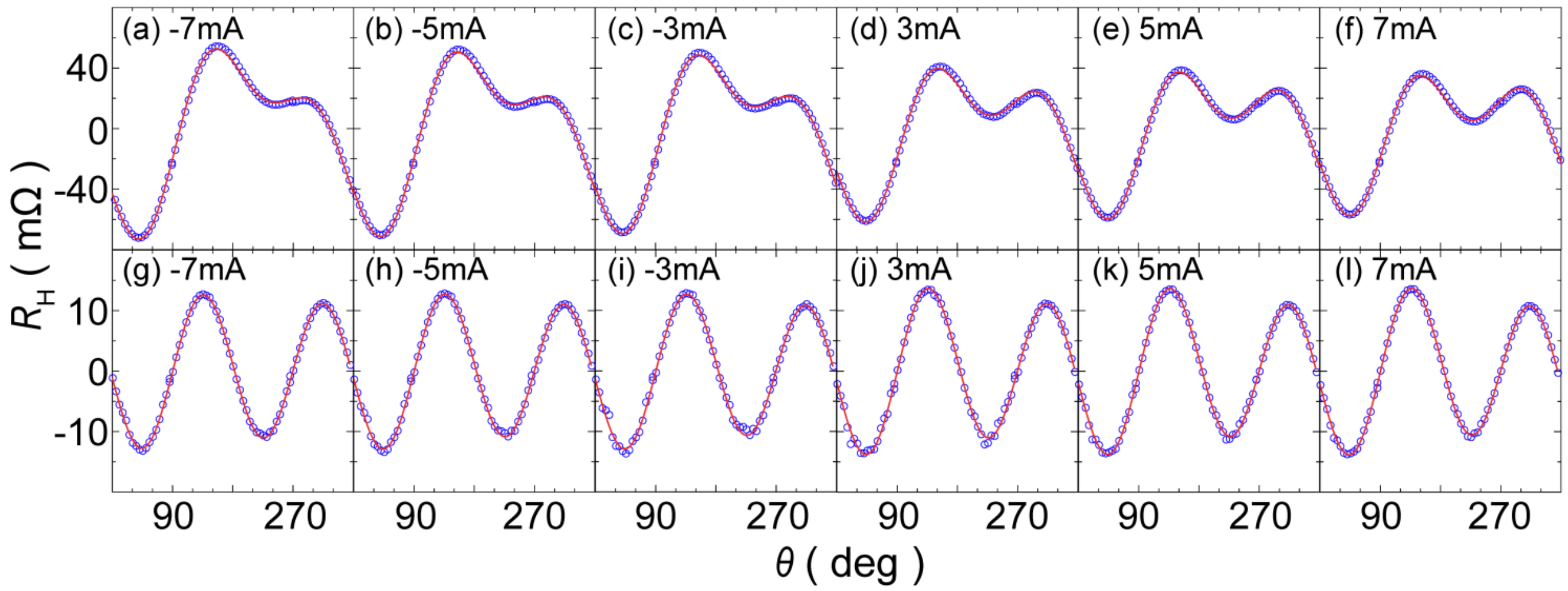

**FIG. 2.** In-plane magnetic field direction ($\theta$)–dependence of the Hall resistance $R_H$. **(a) – (f)** sample A at 8 K, and **(g) – (l)** sample B at 10 K. The bias current is changed from -7 mA to 7 mA. Blue circles are experiment data and red lines are fitting using Eq. (2) in text.

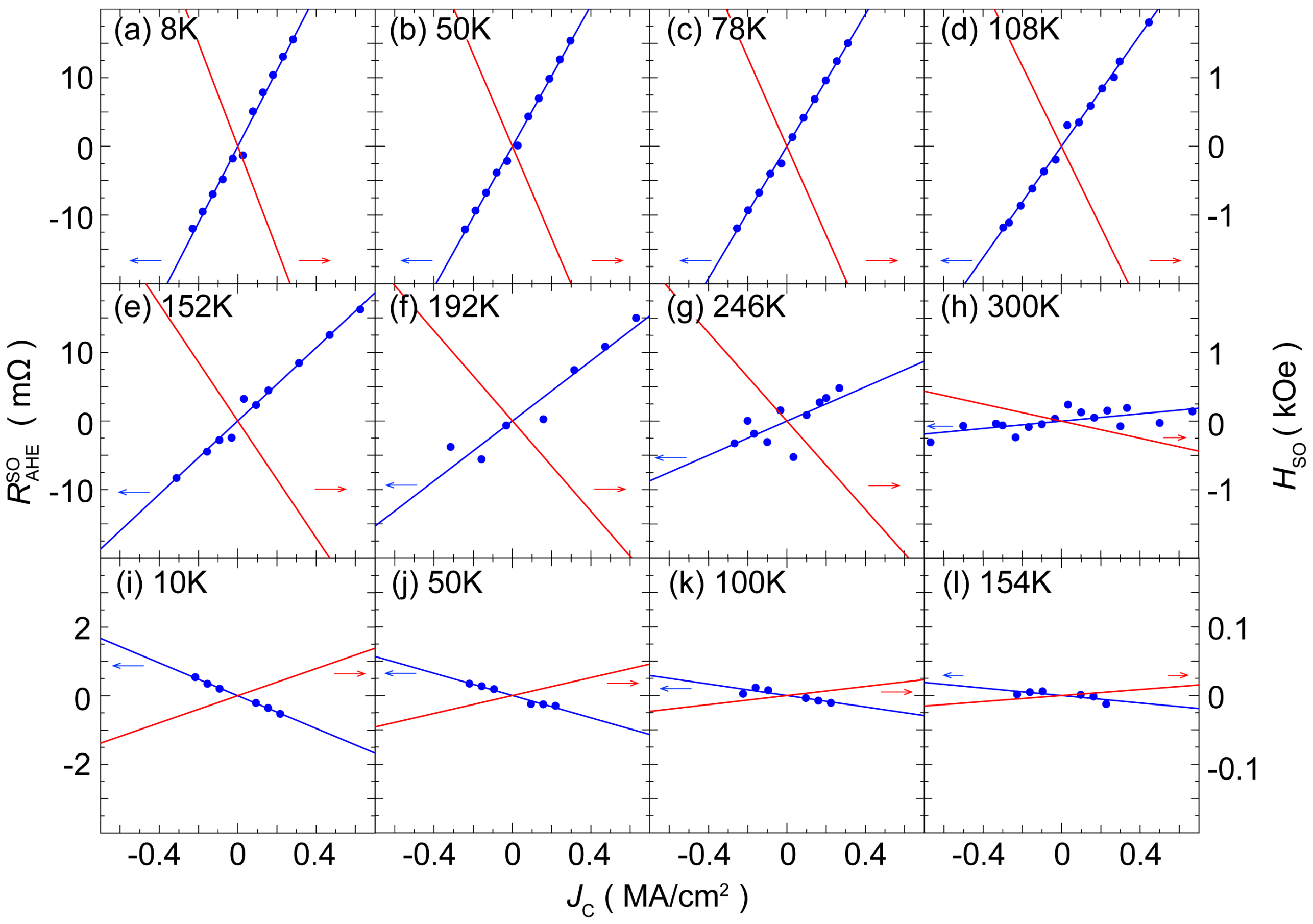

**FIG. 3.** Amplitude of the anomalous Hall resistance $R_{\mathrm{AHE}}^{\mathrm{SO}}$ arising from the spin-orbit effective field $H_{\mathrm{SO}}$, as a function of the charge current density in the BiSb layer at various temperatures for **(a)-(h)** sample A and **(i)-(l)** sample B. Blue dots and blue lines are the experimental data and linear fitting of $R_{\mathrm{AHE}}^{\mathrm{SO}}$, while red solid lines show the calculated $H_{\mathrm{SO}}$, respectively.

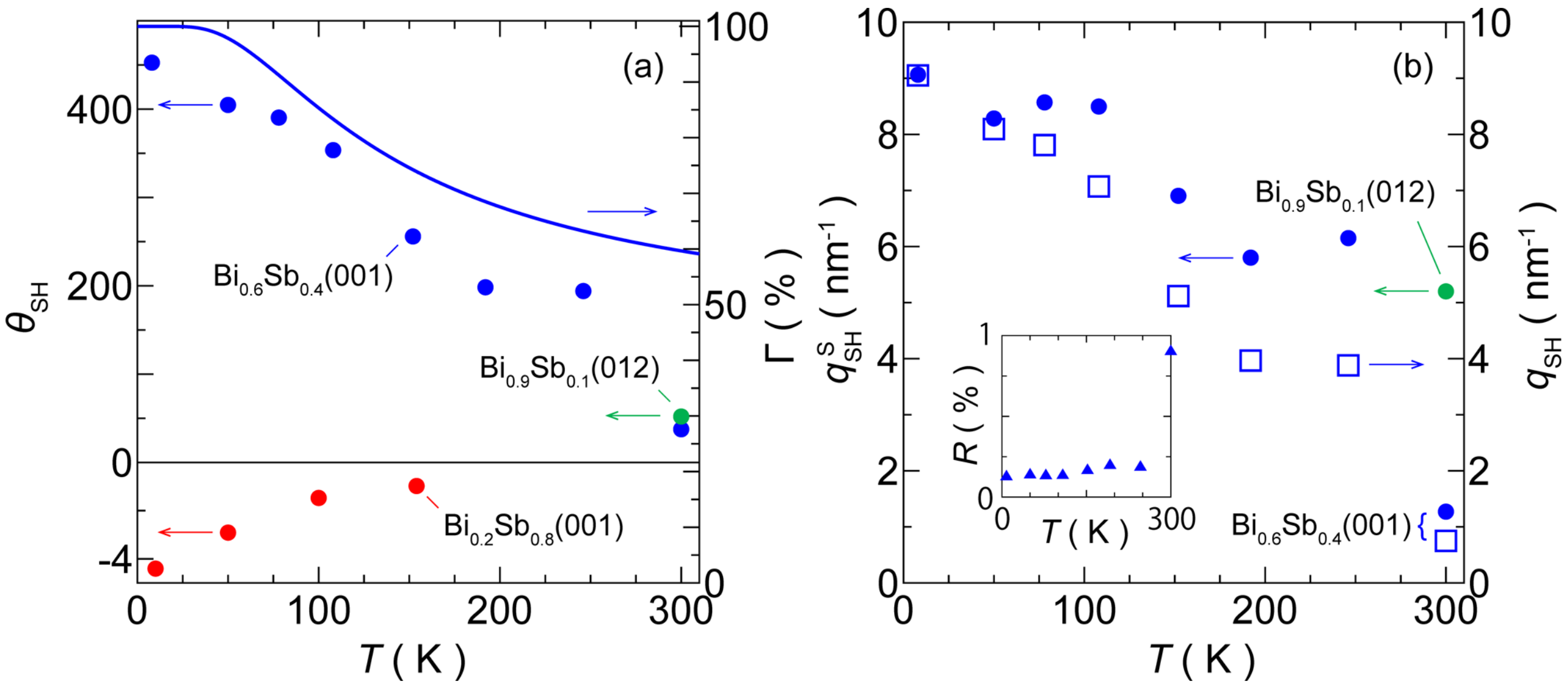

**FIG. 4. (a)** Temperature dependence of the nominal spin Hall angle $\theta_{\mathrm{SH}}$ of the 50 nm-thick $Bi_{0.6}Sb_{0.4}(001)$ layer in sample A (blue dots) and the 50 nm-thick $Bi_{0.2}Sb_{0.8}(001)$ layer in sample B (red dots). Blue line shows the surface contribution factor $\Gamma$ to the total conductivity in the $Bi_{0.6}Sb_{0.4}(001)$ layer. **(b)** Temperature dependence of the nominal sheet spin Hall angle $q_{\mathrm{SH}}$ (blue squares) and the surface sheet spin Hall angle $q_{\mathrm{SH}}^{\mathrm{S}}$ (blue dots) of sample A. Inset shows the temperature dependence of the bulk contribution to the nominal sheet spin Hall angle, $R \equiv \frac{2e}{\hbar}\frac{\left|\sigma_{\mathrm{SH}}^{\mathrm{B}}\right|}{q_{\mathrm{SH}}(\sigma_{\mathrm{S}} t_{\mathrm{S}} + \sigma_{\mathrm{B}} t_{\mathrm{B}})}$. For a reference, we also show the room-temperature $\theta_{\mathrm{SH}}$ and $q_{\mathrm{SH}}^{\mathrm{S}}$ of a 10 nm-thick $Bi_{0.9}Sb_{0.1}(012)$ layer (green dots in **(a)** and **(b)**).

**Supplementary Information**

# Origin of the giant spin Hall effect in BiSb topological insulator

Takanori Shirokura[1], Kenichiro Yao[1], Yugo Ueda[1], and Pham Nam Hai[1,2,3*]

[1]*Department of Electrical and Electronic Engineering, Tokyo Institute of Technology,*

*2-12-1 Ookayama, Meguro, Tokyo 152-0033, Japan*

[2]*Center for Spintronics Research Network (CSRN), The University of Tokyo,*

*7-3-1 Hongo, Bunkyo, Tokyo 113-8656, Japan*

[3]*CREST, Japan Science and Technology Agency,*

*4-1-8 Honcho, Kawaguchi, Saitama 332-0012, Japan*

*Corresponding author: pham.n.ab@m.titech.ac.jp

## Note 1: Temperature dependence of the resistivity of $Bi_{0.4}Sb_{0.6}$ and $Bi_{0.2}Sb_{0.8}$

Figure S1 shows the temperature dependence of the resistivity of a $Bi_{0.4}Sb_{0.6}$ and a $Bi_{0.2}Sb_{0.8}$ single layer (50 nm). The solid line is the fitting curve using Eq. (1) in the main text, which shows that the $Bi_{0.6}Sb_{0.4}$ layer behaves as a topological insulator with a band gap of 33.2 meV and mixing of surface and bulk conduction, and that Γ changes from 60 % at 300 K to 100 % at 8 K. Meanwhile, the $Bi_{0.2}Sb_{0.8}$ layer behaves purely as a semimetal layer, whose resistivity monotonically decreases with lowering temperature.

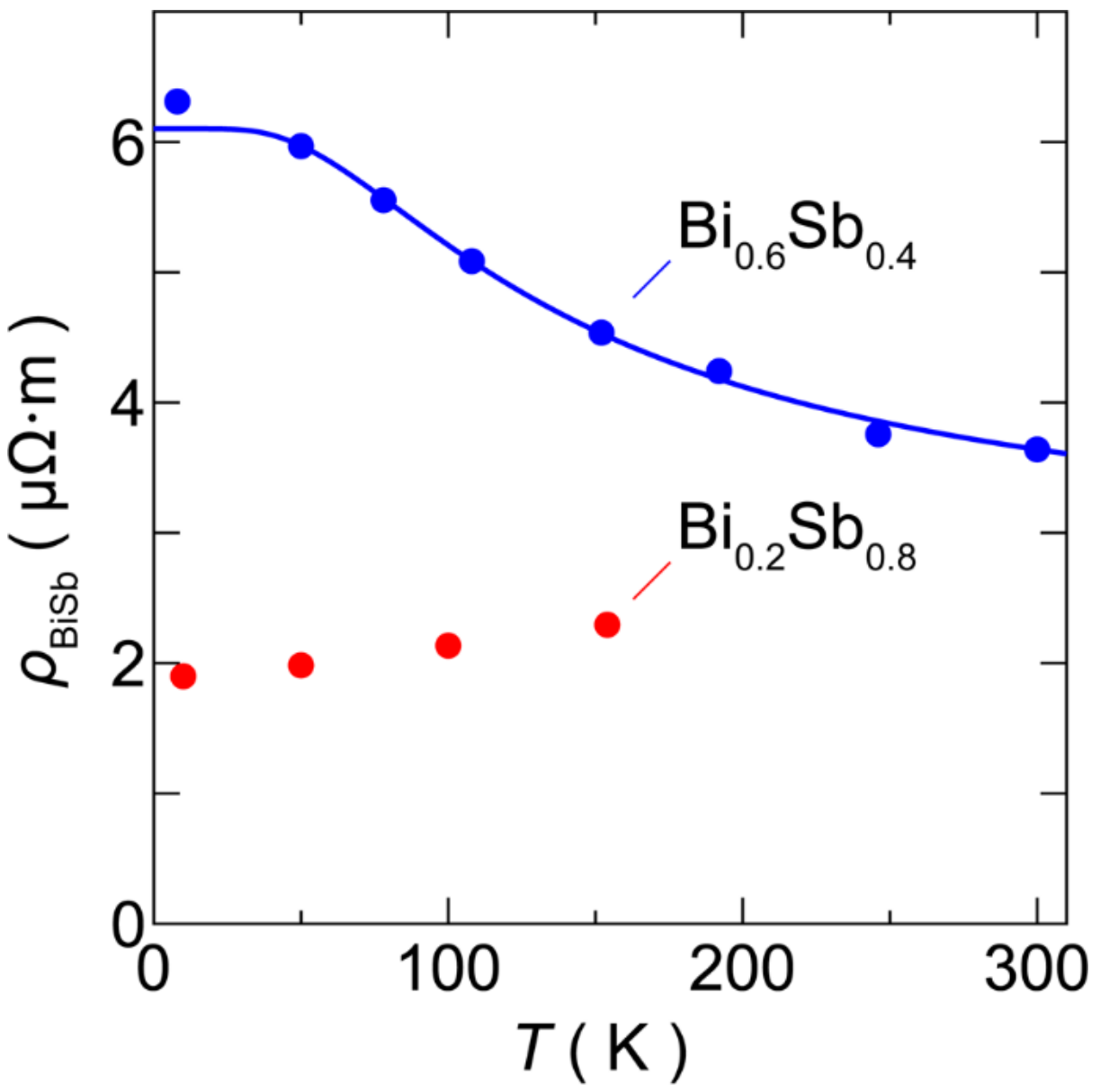

**Fig. S1.** Temperature dependence of the resistivity of a $Bi_{0.4}Sb_{0.6}$ (blue dots) and a $Bi_{0.2}Sb_{0.8}$ (red dots) single layer (50 nm). The blue line is the fitting curve using Eq. (1) in the main text.

## Note 2: Estimation of the conductivity of the top MnAs layers and the spin Hall angle from the planar Hall resistance

In this work, the MnAs layers are grown on top of the BiSb layers. Thus, its conductivity can be different from that of a single crystalline stand-alone MnAs thin film, and it is not appropriate to use the conductivity of the single crystalline MnAs layer for estimation of the current density in the MnAs and BiSb layers. Here, we estimated the conductivity of the top MnAs layers by comparing the amplitude of their planar Hall resistance to that of a standard 23.3 nm-thick MnAs layer (the "standard" MnAs) grown on a GaAs(111)B substrate, whose planar Hall resistance as a function of its conductivity was measured in advance at various temperatures. Taking into account the shunt circuit effect of the bias current and short circuit effect of the Hall voltage [1], the planar Hall resistance $R_{\mathrm{PHE}}(\phi)$ of the standard MnAs single layer and that of the MnAs/BiSb bi-layers are given by

$$R_{\mathrm{PHE}}(\phi) = \frac{A(\sigma_{\mathrm{MnAs}})}{t_{\mathrm{MnAs}}} \sin 2\phi, \quad \text{(single layer)} \tag{S1}$$

$$R_{\mathrm{PHE}}(\phi) = \frac{A(\sigma_{\mathrm{MnAs}})}{t_{\mathrm{MnAs}}} \left( \frac{\sigma_{\mathrm{MnAs}} t_{\mathrm{MnAs}}}{\sigma_{\mathrm{MnAs}} t_{\mathrm{MnAs}} + \sigma_{\mathrm{BiSb}} t_{\mathrm{BiSb}}} \right)^2 \sin^2 \psi \sin 2\phi \text{ , (bi − layers)} \tag{S2}$$

where $A = \frac{1}{2}(\rho_{\perp} - \rho_{\parallel})$ reflects the magnitude of the planar Hall resistivity and is given by half of the difference between the transverse $(\rho_{\perp})$ and longitudinal $(\rho_{\parallel})$ magnetoresistivity [2], $\sigma_{\mathrm{MnAs}}$ and $t_{\mathrm{MnAs}}$ are the conductivity and the thickness of MnAs, $\sigma_{\mathrm{BiSb}}$ and $t_{\mathrm{BiSb}}$ are the conductivity and the thickness of BiSb, $\phi$ is the azimuth angle of the magnetization with respect to the charge current direction, and $\psi$ is the polar angle of the magnetization, which is attributed to the z-component of the magnetization arising from the spin-orbit torque. In Eq. (S1) and (S2), $A$ is a function of the conductivity of MnAs, and can be deduced in advance by measuring $R_{\mathrm{PHE}}$ and $\sigma_{\mathrm{MnAs}}$ of the single MnAs layer at various temperatures. Thus, we can estimate $\sigma_{\mathrm{MnAs}}$ in the bi-layers by measuring $R_{\mathrm{PHE}}$ as shown in Eq. (S2).

By solving the torque balance equation involving the in-plane external field $H_{\mathrm{ext}}$, the spin orbit effective field $H_{\mathrm{SO}}$, and the demagnetising field $H_{\mathrm{deg}} = 4\pi M_{\mathrm{S}}$, we obtain

$$\cos \psi = \frac{H_{\mathrm{so}} \cos\phi}{(H_{\mathrm{ext}} + H_{\mathrm{deg}})} .$$

Thus,

$$\begin{aligned} R_{\mathrm{PHE}}(\phi) &= \frac{A}{t_{\mathrm{MnAs}}} \left( \frac{\sigma_{\mathrm{MnAs}} t_{\mathrm{MnAs}}}{\sigma_{\mathrm{MnAs}} t_{\mathrm{MnAs}} + \sigma_{\mathrm{BiSb}} t_{\mathrm{BiSb}}} \right)^2 \left[ 1 - \frac{(H_{\mathrm{so}} \cos\phi)^2}{(H_{\mathrm{ext}} + H_{\mathrm{deg}})^2} \right] \sin 2\phi \\ &= \frac{A}{t_{\mathrm{MnAs}}} \left( \frac{\sigma_{\mathrm{MnAs}} t_{\mathrm{MnAs}}}{\sigma_{\mathrm{MnAs}} t_{\mathrm{MnAs}} + \sigma_{\mathrm{BiSb}} t_{\mathrm{BiSb}}} \right)^2 \left\{ \left[ 1 - \frac{(H_{\mathrm{so}})^2}{2(H_{\mathrm{ext}} + H_{\mathrm{deg}})^2} \right] \sin 2\phi \right. \\ &\qquad \left. - \frac{(H_{\mathrm{so}})^2}{4(H_{\mathrm{ext}} + H_{\mathrm{deg}})^2} \sin 4\phi \right\} \end{aligned}$$

The $\sin 4\phi$ term in the bracket is small compared with the $\sin 2\phi$ term and can be dropped. We arrive at

$$R_{\mathrm{PHE}}(\phi) = \frac{A}{t_{\mathrm{MnAs}}}\left(\frac{\sigma_{\mathrm{MnAs}}t_{\mathrm{MnAs}}}{\sigma_{\mathrm{MnAs}}t_{\mathrm{MnAs}} + \sigma_{\mathrm{BiSb}}t_{\mathrm{BiSb}}}\right)^2\left[1 - \frac{(H_{\mathrm{so}})^2}{2(H_{\mathrm{ext}} + H_{\mathrm{deg}})^2}\right]\sin 2\phi. \quad \text{(S3)}$$

Eq. (S3) shows that the amplitude of $R_{\mathrm{PHE}}^{\mathrm{B}}(\phi)$ (denoted below by $R_{\mathrm{PHE}}^{\mathrm{B}}$) is a quadratic function of $H_{\mathrm{SO}}$:

$$\begin{aligned} R_{\mathrm{PHE}} &= \frac{|A|}{t_{\mathrm{MnAs}}}\left(\frac{\sigma_{\mathrm{MnAs}}t_{\mathrm{MnAs}}}{\sigma_{\mathrm{MnAs}}t_{\mathrm{MnAs}} + \sigma_{\mathrm{BiSb}}t_{\mathrm{BiSb}}}\right)^2\left[1 - \frac{(H_{\mathrm{SO}})^2}{2(H_{\mathrm{ext}} + H_{\mathrm{deg}})^2}\right] \\ &= R_{\mathrm{PHE-0}}\left[1 - \frac{(H_{\mathrm{SO}})^2}{2(H_{\mathrm{ext}} + H_{\mathrm{deg}})^2}\right] \qquad \text{(S4)} \end{aligned}$$

$$R_{\mathrm{PHE-0}} \equiv \frac{|A|}{t_{\mathrm{MnAs}}}\left(\frac{\sigma_{\mathrm{MnAs}}t_{\mathrm{MnAs}}}{\sigma_{\mathrm{MnAs}}t_{\mathrm{MnAs}} + \sigma_{\mathrm{BiSb}}t_{\mathrm{BiSb}}}\right)^2 \qquad \text{(S5)}$$

Eq. (S4) has two very important applications. First, it indicates that $R_{\mathrm{PHE}}$ is reduced by

$$\frac{\Delta R_{\mathrm{PHE}}}{R_{\mathrm{PHE}}} = -\frac{(H_{\mathrm{SO}})^2}{2(H_{\mathrm{ext}} + H_{\mathrm{deg}})^2} = -\frac{(\alpha I)^2}{2(H_{\mathrm{ext}} + H_{\mathrm{deg}})^2}, \text{(S6)}$$

where $\alpha$ is a proportional constant related to the spin Hall angle, and $I$ is the bias charge current. In heavy metals, $H_{\mathrm{SO}}$ is typically small (~ 10 Oe·MA$^{-1}$cm$^{-2}$) compared with $H_{\mathrm{deg}}$, thus the change of $R_{\mathrm{PHE}}$ is undetectable. In our case of BiSb, $H_{\mathrm{SO}}$ is as large as a few kOe·MA$^{-1}$cm$^{-2}$ while $H_{\mathrm{ext}} + H_{\mathrm{deg}}$ = 10 kOe, thus the change of $R_{\mathrm{PHE}}$ is detectable. Therefore, $H_{\mathrm{SO}}$ can be determined with high certainty by fitting the $\frac{\Delta R_{\mathrm{PHE}}}{R_{\mathrm{PHE}}}$ - $I$ data to Eq. (S6) without the need to know the exact value of $\sigma_{\mathrm{MnAs}}$ and $\sigma_{\mathrm{BiSb}}$ in the bi-layer (although their exact values can be determined as follows).

To determine $\sigma_{\mathrm{MnAs}}$, we rewrite Eq. (S5) as

$$R_{\mathrm{PHE-0}} = |A|t_{\mathrm{MnAs}}(\sigma_{\mathrm{MnAs}})^2\frac{W^2}{(G_{\mathrm{tot}})^2L^2}, \qquad \text{(S7)}$$

where $G_{\mathrm{tot}}, W, L$ are the total conductance, width, and length of the MnAs/BiSb Hall bars. Let $y = \sigma_{\mathrm{MnAs}}$ a variable to be determined in Eq. (S7), we have

$$|A(y)| * y^2 = \frac{L^2R_{\mathrm{PHE-0}}(G_{\mathrm{tot}})^2}{W^2t_{\mathrm{MnAs}}}. \qquad \text{(S8)}$$

Since the function $A(y)$ is known by measuring in advance the relationship between $R_{\mathrm{PHE}}$ and $\sigma_{\mathrm{MnAs}}$ of the standard MnAs single layer in Eq. (S1) at various temperatures, we can deduce $y$ from the experiment data of $R_{\mathrm{PHE-0}}$.

Fig. S2 shows representative $I$-dependence of $R_{\mathrm{PHE}}$ in sample A and B. The blue dots show the measurement values and the blue solid curves show the fitting to Eq. (S6). Clear quadratic dependence of $R_{\mathrm{PHE}}$ on $I$ was observed due to the strong $H_{\mathrm{SO}}$ in sample A, as shown in Fig. S2(a) – S2(d). For example, data in Fig. S2(a) yield $\alpha = 0.22$ kOe/mA and $\theta_{\mathrm{SH}}$ = 530 at 8 K. However, no clear change of $R_{\mathrm{PHE}}$ with $I$ was observed in sample B, in agreement with the smaller SHE in sample B.

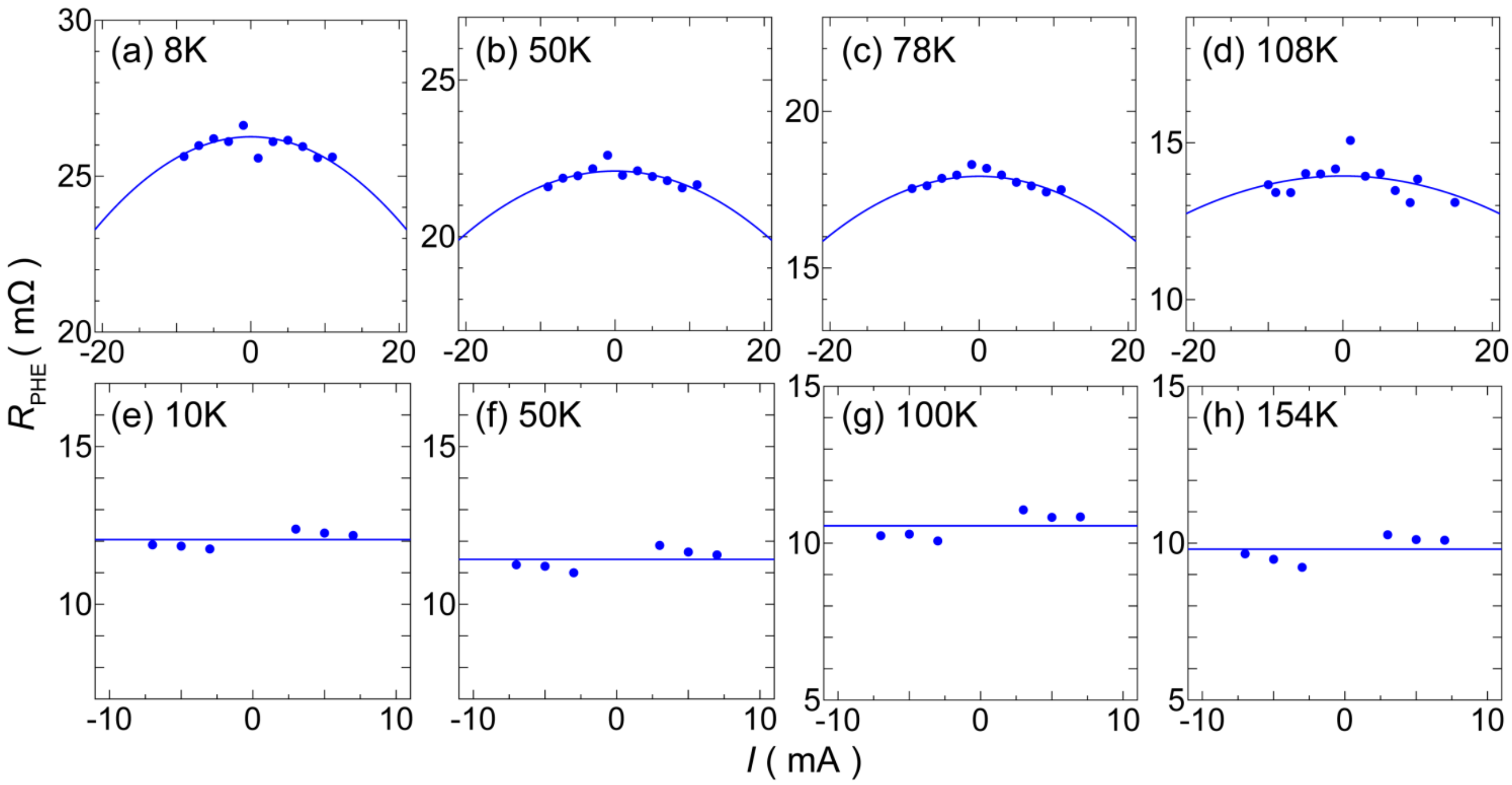

**Fig. S2.** $I$-dependence of the planar Hall resistance $R_{\mathrm{PHE}}$ for **(a)** – **(d)** sample A and **(e)** – **(h)** sample B.

Fig. S3(a) shows the relationship between $R_{\mathrm{PHE}}$ and $\sigma_{\mathrm{BiSb}}$ of the standard MnAs single layer, which is then used to estimate $\sigma_{\mathrm{MnAs}}$ for sample A and B. The blue dots show the measurement values and the blue solid line shows a fitting curve using a fourth degree polynomial function. $\sigma_{\mathrm{MnAs}}$ is then estimated by solving Eq. (S8) with $R_{\mathrm{PHE-0}}$ obtained from the $I$-dependence data of $R_{\mathrm{PHE}}$ at the $I \rightarrow 0$ mA limit. Fig. S3(b) shows the temperature dependence of $\sigma_{\mathrm{MnAs}}$ in sample A and B, which are then used to estimate $J_{\mathrm{C}}$ and $\theta_{\mathrm{SH}}$.

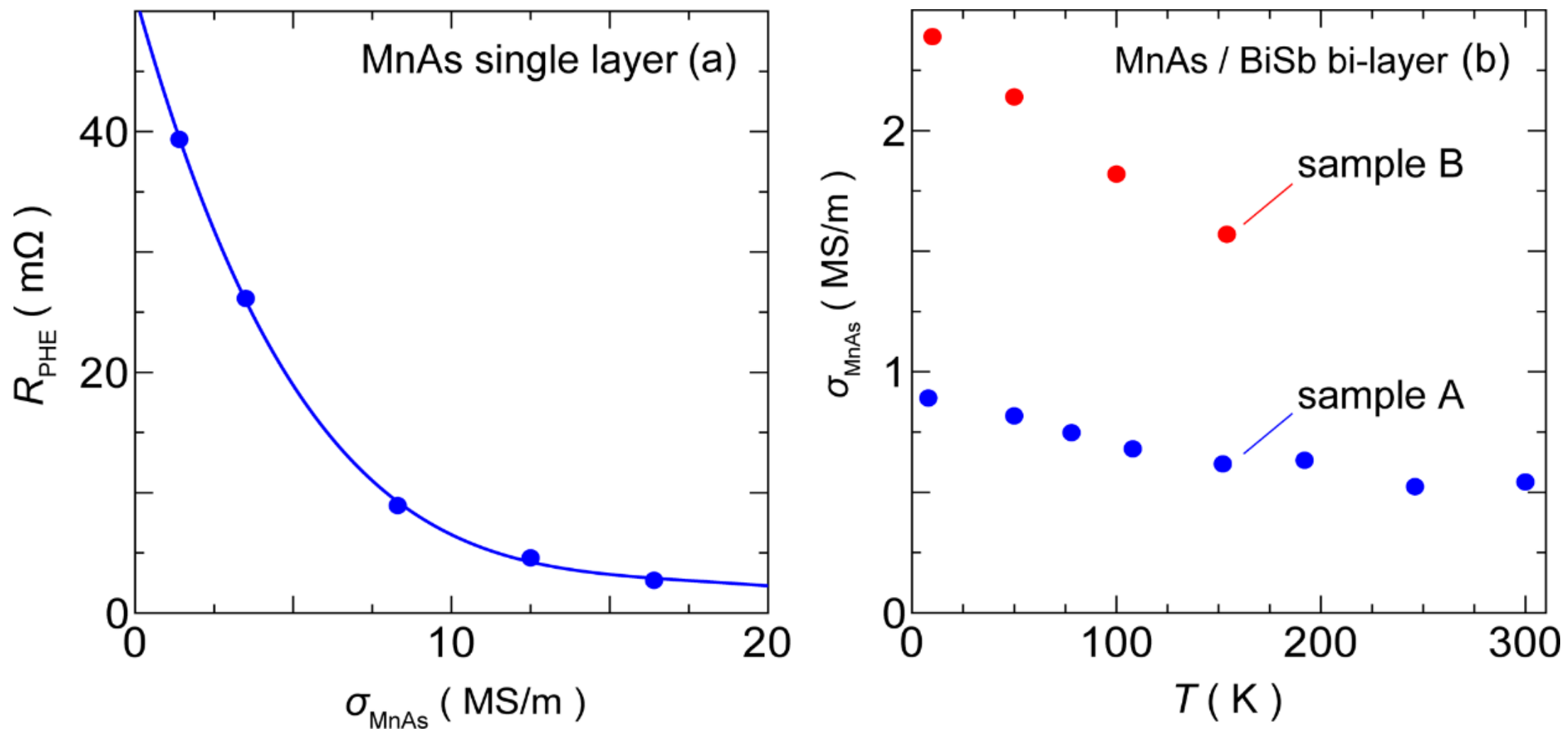

**Fig. S3.** (a) Relationship between the planar Hall resistance $R_{\mathrm{PHE}}$ and the conductivity $\sigma_{\mathrm{MnAs}}$ in a 23.3 nm-thick MnAs single layer. (b) Temperature dependence of the conductivity $\sigma_{\mathrm{MnAs}}$ of the MnAs top layer in the MnAs/BiSb bi-layers.

## Note 3: Measurements of d$R_{\mathrm{AHE}}$/d$H_{\mathrm{perp}}$

In order to estimate $H_{\mathrm{SO}}$ from the $R_{\mathrm{H}}^{\mathrm{SO}}$ data, we need to know the anomalous Hall resistance-magnetic field gradient values at each temperature. This was done by measuring the Hall resistance with a perpendicular magnetic field $H_{\mathrm{perp}}$. In our MnAs/BiSb bi-layer systems, the Hall voltage due to $H_{\mathrm{perp}}$ originates from the ordinary Hall voltage from the BiSb layer and the anomalous Hall voltage from the MnAs layer. By considering the shunt circuit effect of the bias current and the short circuit effect of both the Hall voltages [1], the nominal Hall resistance gradient $\frac{\mathrm{d}R_{\mathrm{H}}}{\mathrm{d}H_{\mathrm{perp}}}$ is given by

$$\frac{\mathrm{d}R_{\mathrm{H}}}{\mathrm{d}H_{\mathrm{perp}}} = \frac{\mathrm{d}R_{\mathrm{OHE}}}{\mathrm{d}H_{\mathrm{perp}}} + \frac{\mathrm{d}R_{\mathrm{AHE}}}{\mathrm{d}H_{\mathrm{perp}}}$$

$$\frac{\mathrm{d}R_{\mathrm{OHE}}}{\mathrm{d}H_{\mathrm{perp}}} = \left(\frac{\sigma_{\mathrm{BiSb}}t_{\mathrm{BiSb}}}{\sigma_{\mathrm{MnAs}}t_{\mathrm{MnAs}} + \sigma_{\mathrm{BiSb}}t_{\mathrm{BiSb}}}\right)^2 \left(\frac{\mathrm{d}R_{\mathrm{OHE}}}{\mathrm{d}H_{\mathrm{perp}}}\right)_0$$

$$\frac{\mathrm{d}R_{\mathrm{AHE}}}{\mathrm{d}H_{\mathrm{perp}}} = \left(\frac{\sigma_{\mathrm{MnAs}}t_{\mathrm{MnAs}}}{\sigma_{\mathrm{MnAs}}t_{\mathrm{MnAs}} + \sigma_{\mathrm{BiSb}}t_{\mathrm{BiSb}}}\right)^2 \left(\frac{\mathrm{d}R_{\mathrm{AHE}}}{\mathrm{d}H_{\mathrm{perp}}}\right)_0$$

where $\left(\frac{\mathrm{d}R_{\mathrm{OHE}}}{\mathrm{d}H_{\mathrm{perp}}}\right)_0$ is the ordinary Hall resistance gradient of the BiSb layer when not in contact with the MnAs layer, and $\left(\frac{\mathrm{d}R_{\mathrm{AHE}}}{\mathrm{d}H_{\mathrm{perp}}}\right)_0$ is the anomalous Hall resistance gradient of the MnAs layer when not in contact with the BiSb layer. Thus, $\frac{\mathrm{d}R_{\mathrm{AHE}}}{\mathrm{d}H_{\mathrm{perp}}}$ can be calculated by subtracting $\frac{\mathrm{d}R_{\mathrm{OHE}}}{\mathrm{d}H_{\mathrm{perp}}}$ measured in a single BiSb layer from the $\frac{\mathrm{d}R_{\mathrm{H}}}{\mathrm{d}H_{\mathrm{perp}}}$ value of the bi-layer. Since the anomalous Hall resistance $R_{\mathrm{H}}^{\mathrm{SO}}$ due to $H_{\mathrm{SO}}$ is given by $R_{\mathrm{H}}^{\mathrm{SO}} = \frac{\mathrm{d}R_{\mathrm{AHE}}}{\mathrm{d}H_{\mathrm{perp}}} H_{\mathrm{SO}}$, we can estimate $H_{\mathrm{SO}}$ by $R_{\mathrm{H}}^{\mathrm{SO}} / \frac{\mathrm{d}R_{\mathrm{AHE}}}{\mathrm{d}H_{\mathrm{perp}}}$.

Table S1 and S2 show the measured values of $\frac{\mathrm{d}R_{\mathrm{H}}}{\mathrm{d}H_{\mathrm{perp}}}$, $\left(\frac{\mathrm{d}R_{\mathrm{OHE}}}{\mathrm{d}H_{\mathrm{perp}}}\right)_0$, $\frac{\mathrm{d}R_{\mathrm{OHE}}}{\mathrm{d}H_{\mathrm{perp}}}$, and $\frac{\mathrm{d}R_{\mathrm{AHE}}}{\mathrm{d}H_{\mathrm{perp}}}$ of sample A and B at various temperature, respectively. $\frac{\mathrm{d}R_{\mathrm{H}}}{\mathrm{d}H_{\mathrm{perp}}}$ were measured by sweeping the external magnetic field perpendicular to the substrate. $\left(\frac{\mathrm{d}R_{\mathrm{OHE}}}{\mathrm{d}H_{\mathrm{perp}}}\right)_0$ were measured in 50 nm-thick single $Bi_{1-x}Sb_x$ ($x = 0.4, 0.8$) layers grown on a GaAs(111)A substrate.

**Table. S1.** Temperature dependence of $\frac{\mathrm{d}R_{\mathrm{H}}}{\mathrm{dH}_{\mathrm{perp}}}$, $\left(\frac{\mathrm{d}R_{\mathrm{OHE}}}{\mathrm{d}H_{\mathrm{perp}}}\right)_0$, $\frac{\mathrm{d}R_{\mathrm{OHE}}}{\mathrm{d}H_{\mathrm{perp}}}$, and $\frac{\mathrm{d}R_{\mathrm{AHE}}}{\mathrm{d}H_{\mathrm{perp}}}$ of sample A

| Sample A | | | | |
|---|---|---|---|---|
| $T$ (K) | $\frac{\mathrm{d}R_{\mathrm{H}}}{\mathrm{d}H_{\mathrm{perp}}}$ (mΩ/kOe) | $\left(\frac{\mathrm{d}R_{\mathrm{OHE}}}{\mathrm{d}H_{\mathrm{perp}}}\right)_0$ (mΩ/kOe) | $\frac{\mathrm{d}R_{\mathrm{OHE}}}{\mathrm{d}H_{\mathrm{perp}}}$ (mΩ/kOe) | $\frac{\mathrm{d}R_{\mathrm{AHE}}}{\mathrm{d}H_{\mathrm{perp}}}$ (mΩ/kOe) |
| 8 | -21.2 | -16.1 | -6.60 | -14.6 |
| 50 | -25.6 | -23.1 | -10.4 | -15.1 |
| 78 | -30.0 | -30.6 | -15.3 | -14.7 |
| 108 | -34.8 | -38.3 | -21.2 | -13.7 |
| 152 | -40.0 | -45.2 | -27.6 | -12.4 |
| 192 | -42.5 | -47.2 | -29.4 | -13.2 |
| 246 | -39.7 | -45.9 | -32.0 | -7.66 |
| 300 | -36.7 | -40.5 | -28.2 | -8.52 |

**Table. S2.** Temperature dependence of $\frac{\mathrm{d}R_{\mathrm{H}}}{\mathrm{d}H_{\mathrm{perp}}}$, $\left(\frac{\mathrm{d}R_{\mathrm{OHE}}}{\mathrm{d}H_{\mathrm{perp}}}\right)_0$, $\frac{\mathrm{d}R_{\mathrm{OHE}}}{\mathrm{d}H_{\mathrm{perp}}}$, and $\frac{\mathrm{d}R_{\mathrm{AHE}}}{\mathrm{d}H_{\mathrm{perp}}}$ of sample B

| Sample B | | | | |
|---|---|---|---|---|
| $T$ (K) | $\frac{\mathrm{d}R_{\mathrm{H}}}{\mathrm{d}H_{\mathrm{perp}}}$ (mΩ/kOe) | $\left(\frac{\mathrm{d}R_{\mathrm{OHE}}}{\mathrm{d}H_{\mathrm{perp}}}\right)_0$ (mΩ/kOe) | $\frac{\mathrm{d}R_{\mathrm{OHE}}}{\mathrm{d}H_{\mathrm{perp}}}$ (mΩ/kOe) | $\frac{\mathrm{d}R_{\mathrm{AHE}}}{\mathrm{d}H_{\mathrm{perp}}}$ (mΩ/kOe) |
| 10 | -27.1 | 29.7 | 17.8 | -44.8 |
| 50 | -27.8 | 29.8 | 18.3 | -46.1 |
| 100 | -28.2 | 29.2 | 18.7 | -46.9 |
| 154 | -27.7 | 27.5 | 18.1 | -45.8 |

## Note 4: Absence of thermoelectric effects

In this note, we present evidences that there is no artifact due to thermoelectric effects that can explain the observed colossal spin Hall angle in sample A. First, we point out that the polarity of $\theta_{\mathrm{SH}}$ changes from positive (sample A) to negative (sample B). Furthermore, it rapidly decreases from 450 to -4.4, although the top layer in sample A and B is the same MnAs material with slightly different thickness. This cannot be explained by thermoelectric effects whose thermal voltage should have the same magnitude and polarity for both samples.

The strongest evidence of the absence of thermoelectric effects in our MnAs/BiSb bi-layers is the bias current-dependence of the amplitude of the planar Hall resistance $R_{\mathrm{PHE}}$. As shown in Eq. (S6) in Note 2, SHE causes the quadratic dependence of $R_{\mathrm{PHE}}$ on $I$. In contrast, thermoelectric effects, such as the planar Nernst effect, should cause the linear dependence of $R_{\mathrm{PHE}}$ on $I$. However, data in Fig. S2(a)-(d) show clear quadratic and no linear dependence of $R_{\mathrm{PHE}}$ on $I$ for sample A. Thus, thermoelectric effects are absent or negligible.

## Note 5: Spin Hall angle in the parallel conduction model

The relationship between the spin current and the charge current is given by

$$J_{\mathrm{S}} = \frac{\hbar}{2e}\theta_{\mathrm{SH}}J_{\mathrm{C}}\,, \tag{S9}$$

where $J_{\mathrm{S}}$ is the spin current density, $\hbar$ is the Dirac's constant, $e$ is the elementary charge, $\theta_{\mathrm{SH}}$ is the nominal spin Hall angle, $J_{\mathrm{C}}$ is the nominal charge current density. In the case of topological insulators, $J_{\mathrm{S}}$ is generated not only by the current following in the bulk states but also in the surface states. Hence,

$$J_{\mathrm{S}} = \frac{\hbar}{2e}\left(\theta_{\mathrm{SH}}^{\mathrm{S}}J_{\mathrm{C}}^{\mathrm{S}} + \theta_{\mathrm{SH}}^{\mathrm{B}}J_{\mathrm{C}}^{\mathrm{B}}\right), \tag{S10}$$

where the symbol S and B represent the surface and bulk contribution, respectively. By considering the parallel conduction model [3] and assuming that the lower surface states not in contact with the MnAs top layer does not contribute to spin current injection, the Eq. (S10) can be

rewritten as

$$J_{\mathrm{S}} = \frac{\hbar}{2e}\left(\frac{\sigma_{\mathrm{S}} t_{\mathrm{B}}}{\sigma_{\mathrm{S}} t_{\mathrm{S}} + \sigma_{\mathrm{B}} t_{\mathrm{B}}}\theta_{\mathrm{SH}}^{\mathrm{S}} + \frac{\sigma_{\mathrm{B}} t_{\mathrm{B}}}{\sigma_{\mathrm{S}} t_{\mathrm{S}} + \sigma_{\mathrm{B}} t_{\mathrm{B}}}\theta_{\mathrm{SH}}^{\mathrm{B}}\right) J_{\mathrm{C}}, \qquad \text{(S11)}$$

By comparing Eq. (S9) and Eq. (S11), we obtain

$$\theta_{\mathrm{SH}} = \frac{\sigma_{\mathrm{S}} t_{\mathrm{B}}}{\sigma_{\mathrm{S}} t_{\mathrm{S}} + \sigma_{\mathrm{B}} t_{\mathrm{B}}}\theta_{\mathrm{SH}}^{\mathrm{S}} + \frac{\sigma_{\mathrm{B}} t_{\mathrm{B}}}{\sigma_{\mathrm{S}} t_{\mathrm{S}} + \sigma_{\mathrm{B}} t_{\mathrm{B}}}\theta_{\mathrm{SH}}^{\mathrm{B}} = \Gamma \frac{t_{\mathrm{B}}}{t_{\mathrm{S}}}\theta_{\mathrm{SH}}^{\mathrm{S}} + \frac{2e}{\hbar}\frac{\sigma_{\mathrm{SH}}^{\mathrm{B}} t_{\mathrm{B}}}{\sigma_{\mathrm{S}} t_{\mathrm{S}} + \sigma_{\mathrm{B}} t_{\mathrm{B}}},$$

where $\Gamma \equiv \sigma_{\mathrm{S}} t_{\mathrm{S}}/(\sigma_{\mathrm{S}} t_{\mathrm{S}} + \sigma_{\mathrm{B}} t_{\mathrm{B}})$ is the contribution of the surface states to the total conductivity, and $\sigma_{\mathrm{SH}}^{\mathrm{B}} \equiv \hbar \sigma_{\mathrm{B}} \theta_{\mathrm{SH}}^{\mathrm{B}}/2e$ is the bulk spin Hall conductivity.